\documentclass[conference]{IEEEtran}
\IEEEoverridecommandlockouts
\usepackage{cite}
\usepackage{amsmath,amssymb,amsfonts}
\usepackage{algorithmic}
\usepackage{graphicx}
\usepackage{textcomp}
\usepackage{xcolor}
\usepackage{multirow}
\usepackage{makecell}
\usepackage{booktabs}
\usepackage{comment}
\def\BibTeX{{\rm B\kern-.05em{\sc i\kern-.025em b}\kern-.08em
    T\kern-.1667em\lower.7ex\hbox{E}\kern-.125emX}}
\begin{document}

\title{Impact of Loss Model Selection on Power Semiconductor Lifetime Prediction in Electric Vehicles\\
}

\author{
	Hongjian~Xia\textsuperscript{1},
	Yi~Zhang\textsuperscript{2,3},
	Dao~Zhou\textsuperscript{2},
    Minyou~Chen\textsuperscript{1},
    Wei Lai\textsuperscript{1},
    Yunhai~Wei\textsuperscript{1},
    and~Huai~Wang\textsuperscript{2}\\
    
    \textsuperscript{1}Department of Electric  Engineering, Chongqing University, Chongqing, China\\
    \textsuperscript{2}AAU Energy, Aalborg University, Aalborg, Denmark\\
    \textsuperscript{3}Swiss Federal Institute of Technology Lausanne (EPFL), Lausanne, Switzerland\\
    hongjian\_xia@cqu.edu.cn; yiz@energy.aau.dk; zda@energy.aau.dk; minyouchen@cqu.edu.cn;\\ laiweicqu@126.com, wyh55043@163.com, hwa@energy.aau.dk
}

\maketitle

\begin{abstract}
Power loss estimation is an indispensable procedure to conduct lifetime prediction for power semiconductor device. The previous studies successfully perform steady-state power loss estimation for different applications, but which may be limited for the electric vehicles (EVs) with high dynamics. Based on two EV standard driving cycle profiles, this paper gives a comparative study of power loss estimation models with two different time resolutions, i.e., the output period average and the switching period average. The correspondingly estimated power losses, thermal profiles, and lifetime clearly pointed out that the widely applied power loss model with the output period average is limited for EV applications, in particular for the highly dynamic driving cycle. The difference in the predicted lifetime can be up to 300 times due to the unreasonable choice the loss model, which calls for the industry attention on the differences of the EVs and the importance of loss model selection in lifetime prediction.

\end{abstract}


\section{Introduction}
The electric vehicle (EV) is getting popular in recent years. Power converters play a key role in the EV systems, however, power semiconductor devices such as insulated-gate bipolar transistors (IGBTs) are one of the most vulnerable components in power converters according to a survey in \cite{Yang2011}. Therefore, it is essential to estimate their lifetimes in EV converters.

\begin{figure}[t!]
	\centering
    \includegraphics[width=80mm]{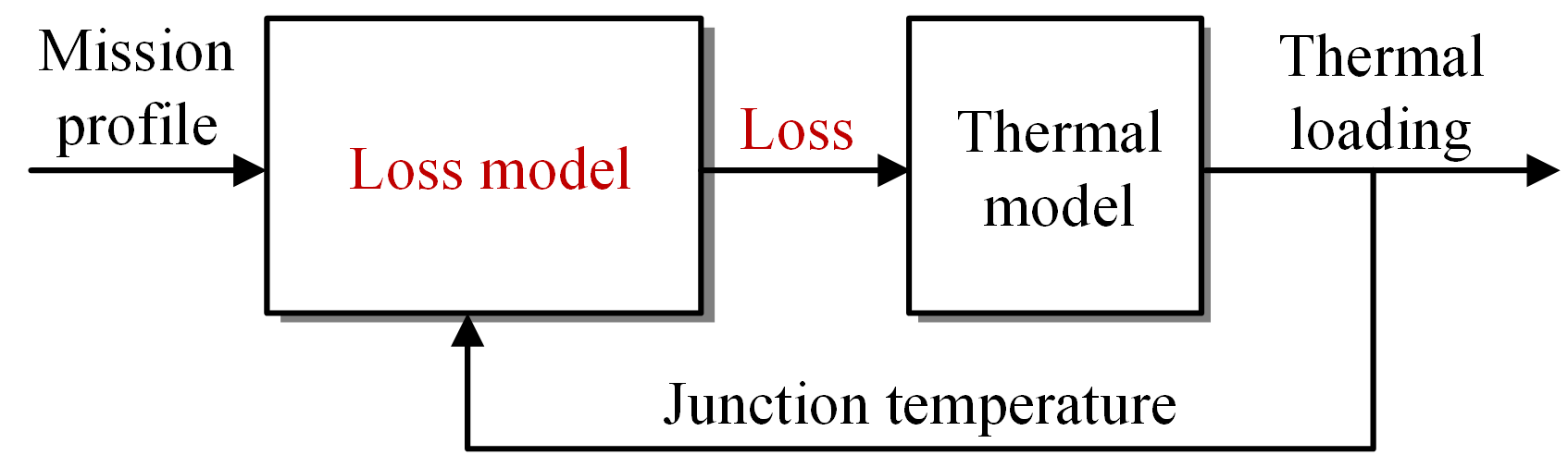}
	\caption{Lifetime estimation flowchart for IGBT module.}
    \label{fig_lifetime prediction process}
\end{figure}

The loss profile translation from the mission profile is one of the important steps in lifetime estimation, as shown in Fig.~\ref{fig_lifetime prediction process}. In previous studies, the loss models are mostly verified in converters that operate in steady state. However, the amplitude and frequency of the output current in EV converters are highly dynamic due to the frequently changed speed and torque. Therefore, the feasibility of applying conventional loss models in the EV converters need to be verified.

The averaging power loss model is firstly developed for two-level converters with the sine pulse width modulation (SPWM) scheme \cite{semiconductors2010application}. The instantaneous power loss on the IGBT is integrated and divided by the fundamental output period to obtain the average power loss. This method is widely applied in grid-connected converters such as wind turbine converters \cite{zhou2020converter}, photovoltaic (PV) converters \cite{Shen2016} and modular multilevel converters \cite{Zhang2019}, etc. However, the main shortcoming of this method is that the power loss over the averaging period is regarded as a constant, which assumes that the corresponding thermal load is a constant. It may be acceptable for the grid-connected converters because the dynamic temperature over the fundamental period is assumed to be negligible \cite{Blasko1999}. For EV converters that operate at dynamic speed and torque, the frequency and amplitude of the output current are changing dynamically. For instance, the output current of the EV converter during acceleration period is in low frequency with large amplitude. Therefore, the thermal fluctuation over the output period could be significantly higher, which should be taken into consideration when estimating the IGBT lifetime in EV converters.

The state-of-the-art have made effort to estimate the IGBT module lifetime in the EV converters. For instance, ref. \cite{Lin2019} estimates the IGBT lifetime with the traction profile in different timescales, and concludes that the long timescale junction temperature fluctuation is the main factor that causes IGBT failure. This conclusion may be reasonable for traction converters because the large junction temperature fluctuation only appears at the adjacent rapid accelerations. However, it is limited to address the reliability challenge of IGBTs in the much more dynamic EV driving cycles. Recently, some researchers use the standard driving cycles to study their influence on the IGBT module lifetime, such as the the urban dynamometer driving schedule (UDDS), the new European driving cycle (NEDC) and the worldwide harmonized vehicles test cycles (WLTC) \cite{Qiu,Sang2017,Zhaksylyk2021,Amirpour2021}, etc. Specifically, ref. \cite{Sang2017} and \cite{Amirpour2021} point out that the driving cycles have a strong impact on the IGBT lifetime. Nevertheless, the aforementioned studies are based on the output period averaging loss model, which neglects the thermal fluctuation during the output period. Unfortunately, the discussion about the impact of loss model averaging time on the lifetime estimation is rarely found. Therefore, further work is still needed to investigate the influence of the loss model averaging time on the IGBT lifetime estimation in EV converter.

The purpose of this paper is to give a comparative study about the impact of loss model selection on the IGBT lifetime estimation in EV converter. Two driving cycle standards, naming the New York city cycle (NYCC) and the highway fuel economy driving schedule (HWFET) are selected as the study case of IGBT lifetime estimation. The outcome of this paper could serve as a reference for the loss model selection for power devices in EV converters under different driving cycles.
\begin{figure}[t!]
	\centering
    \includegraphics[width=80mm]{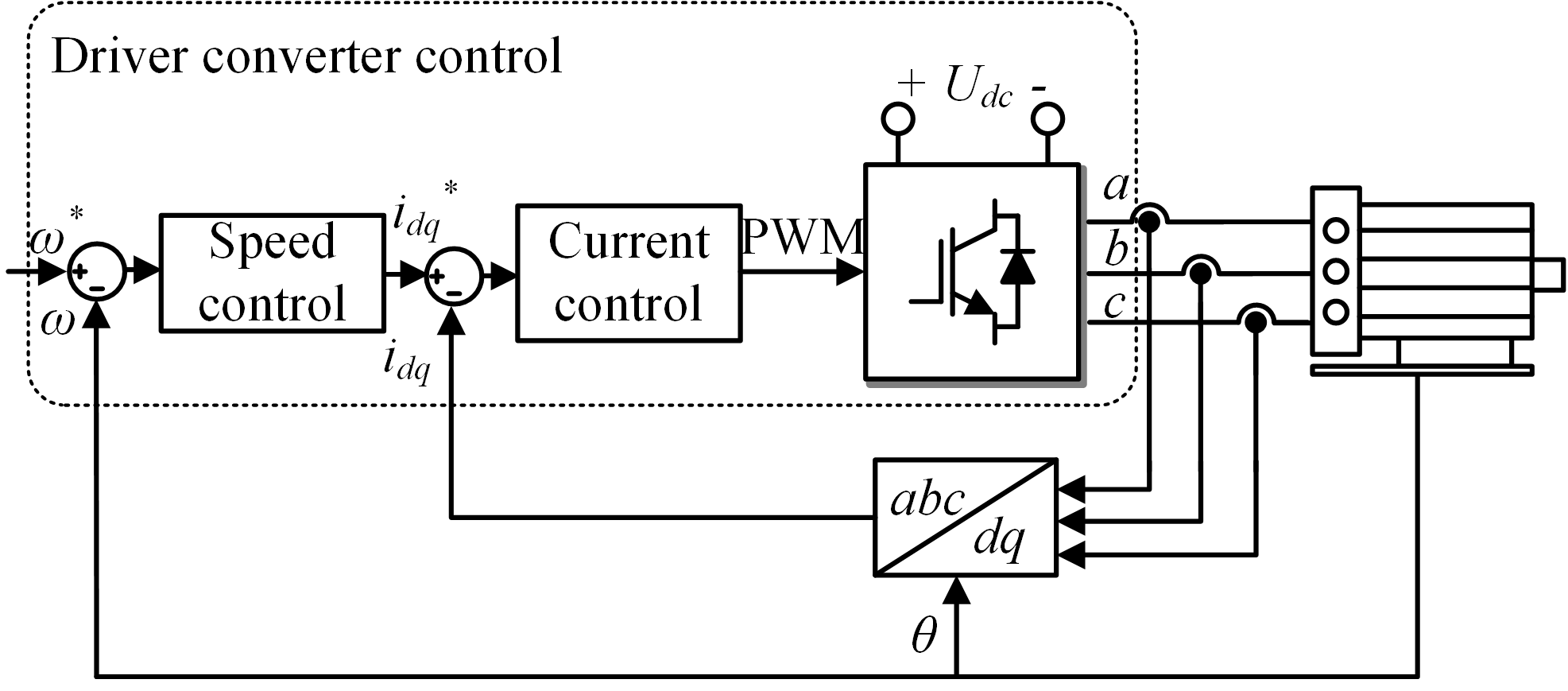}
	\caption{The topology and control scheme of EV converter.}
    \label{fig_PMSM model}
\end{figure}
\section{System description and mission profile in different driving cycles}
\subsection{System configuration}
The EV converter topology is shown in Fig.~\ref{fig_PMSM model}. A three-phase two-level converter is connected to the motor to control the speed and torque. The permanent magnet synchronous motor (PMSM) is chosen as the study case in this paper due to its high efficiency and robustness \cite{Loganayaki2019}. The PMSM is controlled by the $i_d$ = 0 control strategy, and two controllers are used to track the target speed and current. $U_\text{dc}$ is the dc bus voltage. $\omega^*$ and $\omega$ are the target and actual speed, while $i_\text{dq}^*$ and $i_\text{dq}$ are the target and actual current. $\theta$ is the rotor position.

\subsection{Mission profile in different driving cycles}
\begin{figure}[t!]
	\centering
    \includegraphics[width=80mm]{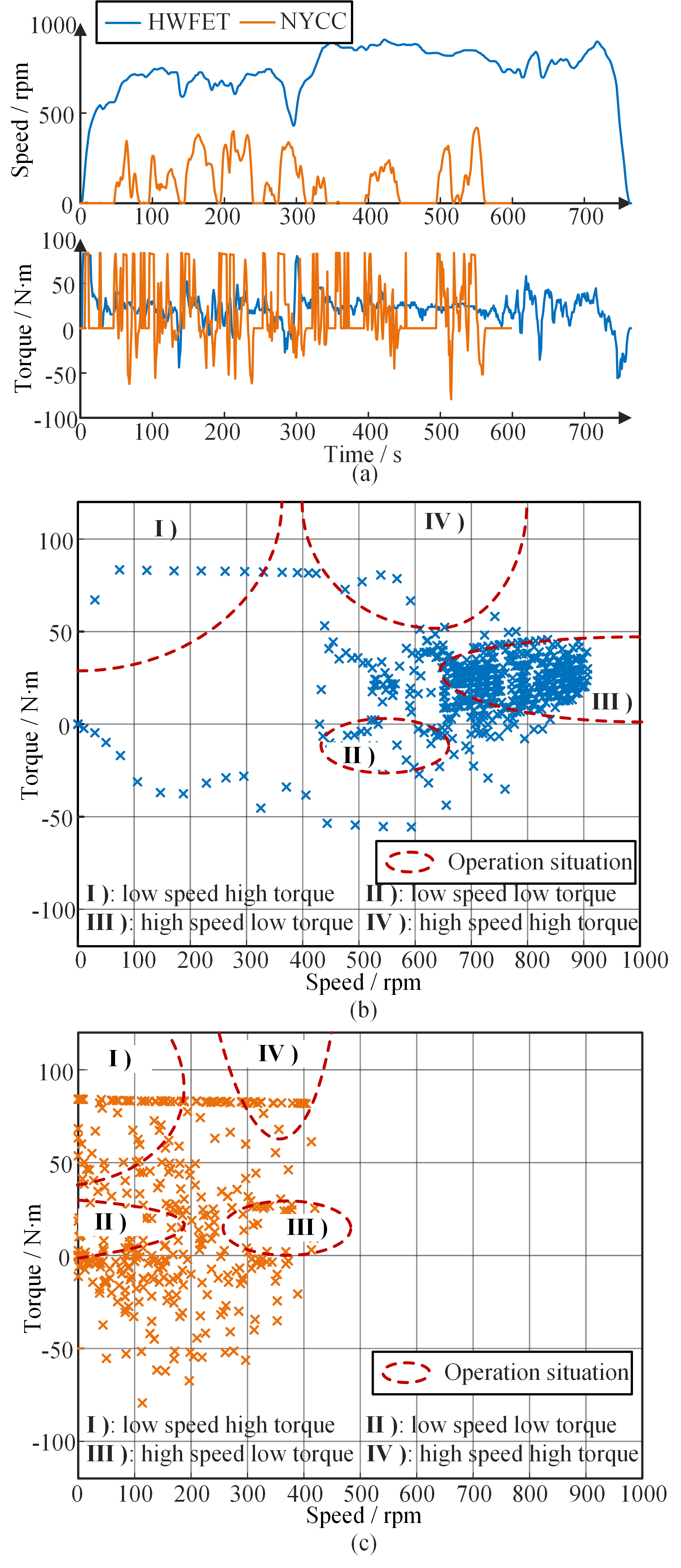}
	\caption{Driving cycle standard of HWFET and NYCC. (a) speed and torque profile. (b) HWFET speed torque map. (c) NYCC speed torque map.}
    \label{fig_driving cycle}
\end{figure}

Different from the conventional two-level converter applications, the EV system has a highly dynamic characteristics because of the average speed and the frequency of stop-and-go. For example, the amplitude and frequency of EV current when the vehicle serving in highway and city are different. Therefore, this section takes two standard driving cycles (NYCC and HWFET) as the study cases, the standard driving cycles and the speed torque maps of HWFET and NYCC are shown in Fig.~\ref{fig_driving cycle}. It can be inferred that the high speed and low torque driving account for the most shares in the HWFET, while the low speed and high torque driving are of most cases in NYCC.

To analyze the difference of converter mission profile under different driving cycles, a PMSM simulation model using the control strategy in Fig.~\ref{fig_PMSM model} is built in PLECS and the system parameters are listed in Table.~\ref{Table: EV emulator parameters}. As shown in Fig.~\ref{fig_driving cycle}(b) and (c), we divide the driving cycles into four different situations to study the difference between the NYCC and HWFET. The four situations have the following performance:
\begin{itemize}
\item[I):]
Low speed and high torque: this situation appears at the acceleration period. In this situation, the current has a high amplitude but low frequency;
\item[II):]
Low speed and low torque: this situation appears when the vehicle is driving at constantly low speed. In this situation, the current has a low amplitude but low frequency;
\item[III):]
High speed and low torque: this situation appears when the vehicle is driving at constantly high speed. In this situation, the current has a low amplitude but high frequency;
\item[IV):]
High speed and high torque: this situation appears at the deceleration period. In this situation, the current has a high amplitude but high frequency.
\end{itemize}

\begin{figure}[t!]
	\centering
    \includegraphics[width=80mm]{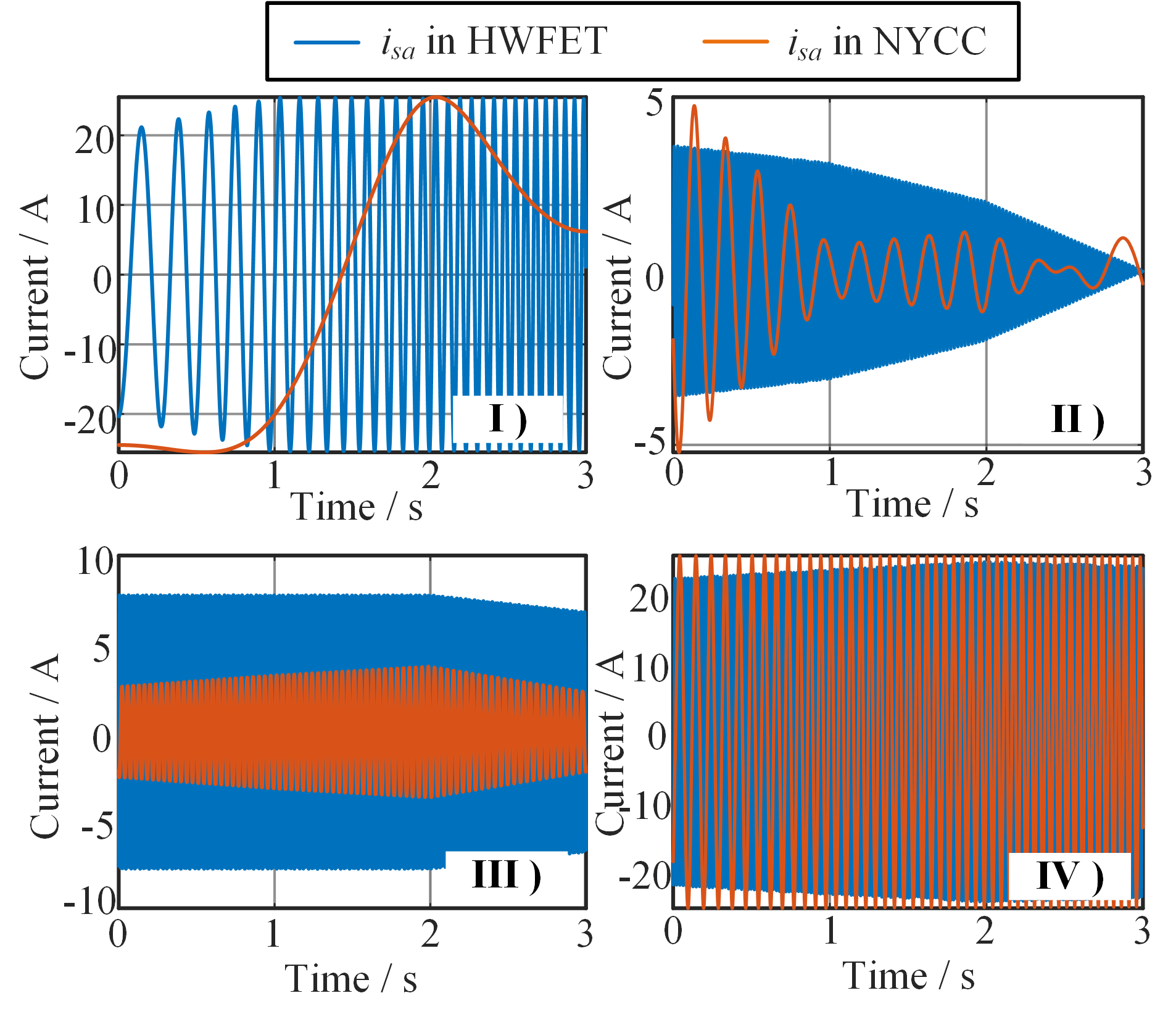}
	\caption{PMSM stator current (phase $a$) compare in different driving cycle.}
    \label{fig_isa compare}
\end{figure}

The output current (take phase $a$ as an example) of the converter in the four situations are shown in Fig.~\ref{fig_isa compare}. It can be inferred from the results that the current amplitude varies with the torque. In addition, the output current frequency under the NYCC is always lower than that under the HWFET due to the lower average speed. Therefore, the mission profile of the EV converter under different driving cycles is highly dynamic, which should be considered in the power loss calculation.
\begin{table}[!t]
	\centering
	\caption{EV emulation parameters}
	\label{Table: EV emulator parameters}
	\begin{tabular}{@{}lcc@{}}
		\toprule
		\textbf{Parameter} &\textbf{Symbols} &\textbf{Value}\\
		\midrule
		DC supply voltage      &$U_\text{dc}$                  &200 V\\
		Switching frequency    &$f_\text{sw}$                  &10 kHz\\
		Stator inductance      &$L_s$                          &5 mH\\
		Stator resistance      &$R_s$                          &0.34 $\Omega$\\
		PM flux linkage        &$\Psi_f$                       &0.022 Wb\\
		Pole pairs             &$p_n$                          &4\\
		Rotational inertia     &$J$                            &0.02 kg$\cdot$m$^2$\\
		Friction coefficient   &$F$                            &0 kg$\cdot$m$^2$/s\\
		\bottomrule
		\end{tabular}
\end{table}

\section{Power loss model verification in different operation situations}
\subsection{Power loss models with different averaging time}
The power dissipation in a power module contains conduction losses and switching losses. The average conduction loss and switching loss in a specific period is
\begin{equation}\label{equ:average conduction loss define}
    P_\text{c} = \frac{1}{t_x}\cdot \int_{0}^{\frac{t_x}{2}} U_{cond}(t)\cdot i(t) dt
\end{equation}
\begin{equation}\label{equ:average switching loss define}
    P_\text{sw} = \frac{1}{t_x}\cdot \Sigma_{n}E_\text{sw}(i(t))
\end{equation}
where $P_c$ is the average conduction power loss, $t_x$ is the averaging period. $U_\text{cond} (t)$ is the conducting voltage and $i(t)$ is the conducting current. $E_\text{sw}$ is the switching energy during the switching.

The output period is usually selected as the averaging period when calculating power loss in converters that mostly operate in steady state, such as the wind turbine converter, PV converter, etc.. In this situation, the averaging conduction loss and switching loss over one output period for IGBT and diode are expressed by
\begin{equation}\label{equ:IGBT conduction loss output}
    P_\text{cS}^\text{$t_o$} = (\frac{1}{2\pi}+\frac{m \text{cos}\varphi}{8}) U_\text{CE} I_m + (\frac{1}{8}+\frac{m \text{cos}\varphi}{3\pi}) r_\text{CE} I_\text{m}^\text{2}
\end{equation}
\begin{equation}\label{equ:Diode conduction loss output}
    P_\text{cD}^\text{$t_o$} = (\frac{1}{2\pi}-\frac{m \text{cos}\varphi}{8}) U_\text{F} I_m + (\frac{1}{8}-\frac{m \text{cos}\varphi}{3\pi}) r_\text{F} I_\text{m}^\text{2}
\end{equation}
\begin{equation}\label{equ:IGBT switching loss output}
    P_\text{swS}^\text{$t_o$} = \frac{1}{\pi}\cdot f_\text{sw} \cdot [E_\text{on}(i(t),U_\text{dc}) + E_\text{off}(i(t),U_\text{dc})] \cdot \frac{I_m}{I^*} \cdot \frac{U_\text{dc}}{U^*}
\end{equation}
\begin{equation}\label{equ:Diode switching loss output}
    P_\text{recD}^\text{$t_o$} = \frac{1}{\pi}\cdot f_\text{sw} \cdot E_\text{rec}(i(t),U_\text{dc}) \cdot \frac{I_m}{I^*} \cdot \frac{U_\text{dc}}{U^*}
\end{equation}
where $t_o$ is the output period. $P_\text{cS}^{t_o}$, $P_\text{cD}^{t_o}$, $P_\text{swS}^{t_o}$ and $P_\text{swD}^{t_o}$ are the average conduction loss and switching of the IGBT and diode over one output period. $U_\text{dc}$ is the dc-bus voltage and $f_\text{sw}$ is the switching frequency. $U^*$ and $I^*$ are testing dc-bus voltage and current in the datasheet provided by the manufacturer. $U_\text{CE}$, $r_\text{CE}$ are the on-state collector-to-emitter voltage and resistance of the IGBT, and $E_\text{on}$, $E_\text{off}$ are the turn-on and turn-off energy of the IGBT. Similarly, $U_\text{F}$ and $r_\text{F}$ are the recovery voltage and resistance of the diode, and $E_\text{rec}$ is the recovery energy of the diode. $I_m$ denotes the amplitude of load current. $m$ is the modulation index and $\varphi$ is the angle between load current and modulation voltage. With the $i_d$ = 0 control strategy in PMSM, they can be calculated by
\begin{equation}\label{equ:iq}
    I_m = i_q = \frac{2\tau_e}{3p_n \cdot \Psi_\text{f}}
\end{equation}
\begin{equation}\label{equ:ud}
    u_d = R_\text{s}i_\text{d}+L_\text{d}\frac{di_d}{dt}-\omega_\text{e}L_\text{q}i_\text{q}
\end{equation}
\begin{equation}\label{equ:uq}
    u_q = R_\text{s}i_\text{q}+L_\text{q}\frac{di_q}{dt}-\omega_\text{e}L_\text{d}i_\text{d}+\omega_\text{e}\Psi_\text{f}
\end{equation}
\begin{equation}\label{equ:m}
    m = \frac{2\sqrt{u_{d}^{2}+u_{q}^{2}}}{U_\text{dc}}
\end{equation}
\begin{equation}\label{equ:phi}
    \varphi = \text{tan}^{-1}(-\frac{u_d}{u_q}) - \text{tan}^{-1}(-\frac{i_d}{i_q})
\end{equation}
where $\tau_e$ is the electromagnetic torque and $p_n$ is the pole pairs. $L_d$ and $L_q$ are the equivalent inductance of the PMSM stator on the $dq$ axis. $R_s$ is the equivalent resistance of the PMSM stator. $\Psi_\text{f}$ is the flux linkage of the PMSM. $u_d$ and $u_q$ are the $dq$ axis voltage, while $i_d$ and $i_q$ are the $dq$ axis current.

The output period averaging calculation considers the power loss as a constant in one output period. It is effective for less dynamic system or long timescale power loss estimation. However, when it comes to highly dynamic system, such as the EV, the power loss dynamic over one output period would be neglected, and the junction temperature swing in one output period would not be considered in the lifetime estimation. Therefore, shorting the averaging time is necessary for the power loss calculation in a high dynamic system. Accordingly, the averaging power loss over one switching period for IGBT and diode are
\begin{equation}\label{equ:IGBT conduction loss switching}
    P_\text{cS}^\text{$t_\text{sw}$} = [U_\text{CE}i(t) + r_\text{CE}i^\text{2}(t)]\cdot \frac{1+m \text{sin}(\omega t + \varphi)}{2}
\end{equation}
\begin{equation}\label{equ:Diode conduction loss switching}
    P_\text{cD}^\text{$t_\text{sw}$} = [U_\text{F}i(t) + r_\text{F}i^\text{2}(t)]\cdot \frac{1-m \text{sin}(\omega t + \varphi)}{2}
\end{equation}
\begin{equation}\label{equ:IGBT switching loss switching}
    P_\text{swS}^\text{$t_\text{sw}$} = f_\text{sw} \cdot [E_\text{on}(i(t),U_\text{dc}) + E_\text{off}(i(t),U_\text{dc})] \cdot \frac{I_m}{I^*} \cdot \frac{U_\text{dc}}{U^*}
\end{equation}
\begin{equation}\label{equ:Diode switching loss switching}
    P_\text{recD}^\text{$t_\text{sw}$} = f_\text{sw} \cdot E_\text{rec}(i(t),U_\text{dc}) \cdot \frac{I_m}{I^*} \cdot \frac{U_\text{dc}}{U^*}
\end{equation}
where $t_{sw}$ is the switching period. $P_\text{cS}^{t_\text{sw}}$, $P_\text{cD}^{t_\text{sw}}$, $P_\text{swS}^{t_\text{sw}}$ and $P_\text{swD}^{t_\text{sw}}$ are the average conduction loss and switching of the IGBT and diode over one switching period. $\omega$ is the angular frequency. $i(t)$ is the instantaneous current at the calculation time, and the current flow through the IGBT and diode in the switching period is considered as a constant.

Compared with the output period averaging model in (\ref{equ:IGBT conduction loss output})~- (\ref{equ:Diode switching loss output}), the switching period averaging model calculates the power loss with a much smaller timescale. Therefore, the detail dynamic power loss is able to be obtained.

\subsection{Loss model verification with steady state operation}
\begin{figure}[t!]
	\centering
    \includegraphics[width=80mm]{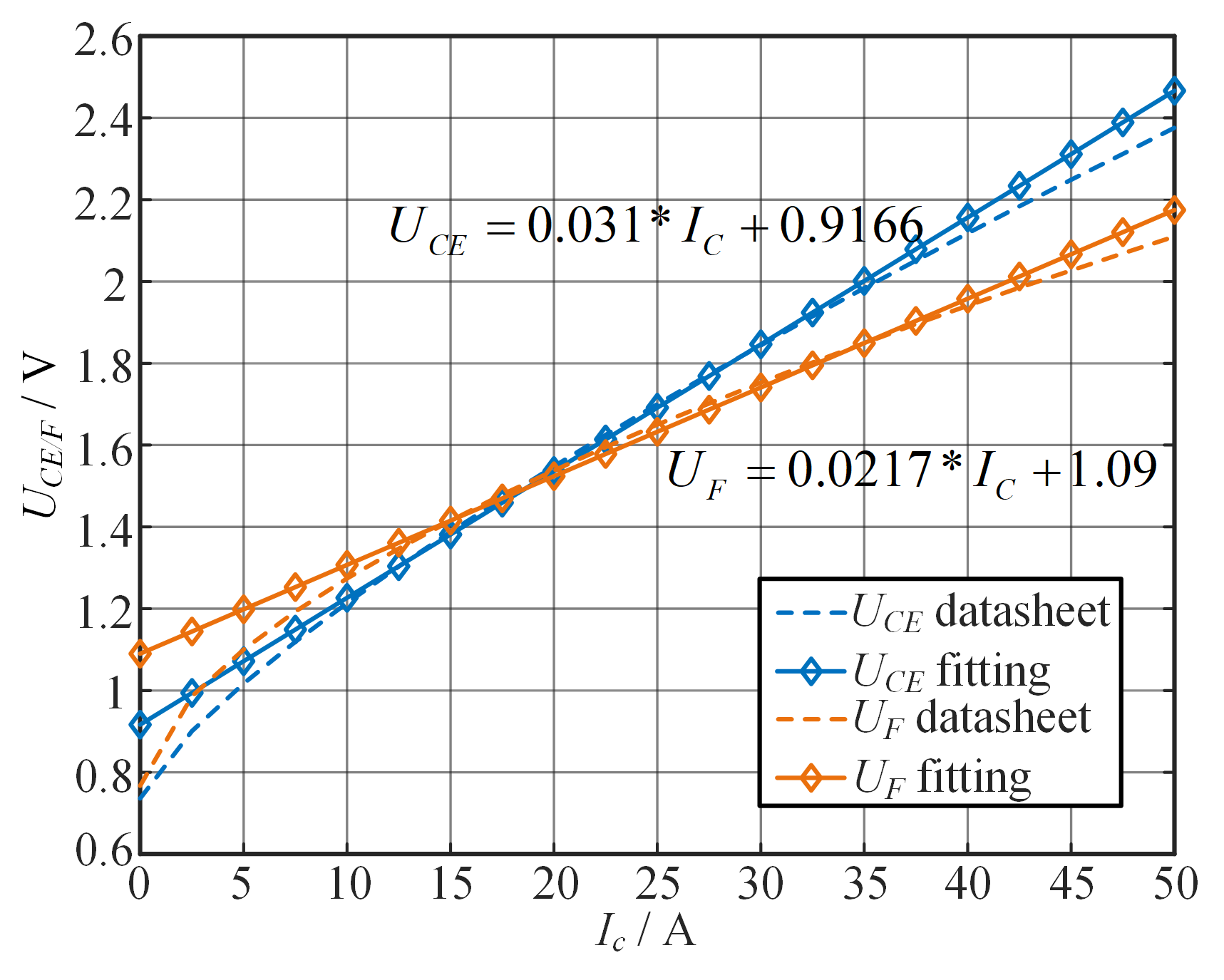}
	\caption{Power device parameters fitting}
    \label{fig_device fitting}
\end{figure}
\begin{table}[!t]
	\centering
	\caption{Downscaled steady state operation points selection}
	\label{Table:Operation points}
	\begin{tabular}{@{}lccccc@{}}
		\toprule
		&\multicolumn{2}{c}{\textbf{HWFET}}&&\multicolumn{2}{c}{\textbf{NYCC}}\\
		\cmidrule{2-3}\cmidrule{5-6}
		&\textbf{Speed}&\textbf{Torque}&&\textbf{Speed}&\textbf{Torque}\\
		&\textbf{(rpm)}&\textbf{(N $\cdot$ m)}&&\textbf{(rpm)}&\textbf{(N $\cdot$ m)}\\
		\midrule
		\textbf{I)}              &362.78           &3.335              &&41.89              &3.367\\
		\textbf{II)}             &362.80           &0.554              &&41.89              &0.556\\
		\textbf{III)}            &906.68           &0.557              &&377.43             &0.752\\
		\textbf{IV)}             &816.18           &2.781              &&377.43             &2.714\\
		\bottomrule
		\end{tabular}
\end{table}
The aforementioned loss models are verified with a benchmark model to prove the feasibility. The benchmark model is built in PLECS with the device thermal description file provided by the manufacturer, the power loss of the power device is obtained by look-up tables in the thermal description file. The power device used in this paper is Infineon FS25R12KT3 module, and the device parameters can be fitted according to the datasheet, which is shown in Fig.~\ref{fig_device fitting}.

\begin{figure}[t!]
	\centering
    \includegraphics[width=80mm]{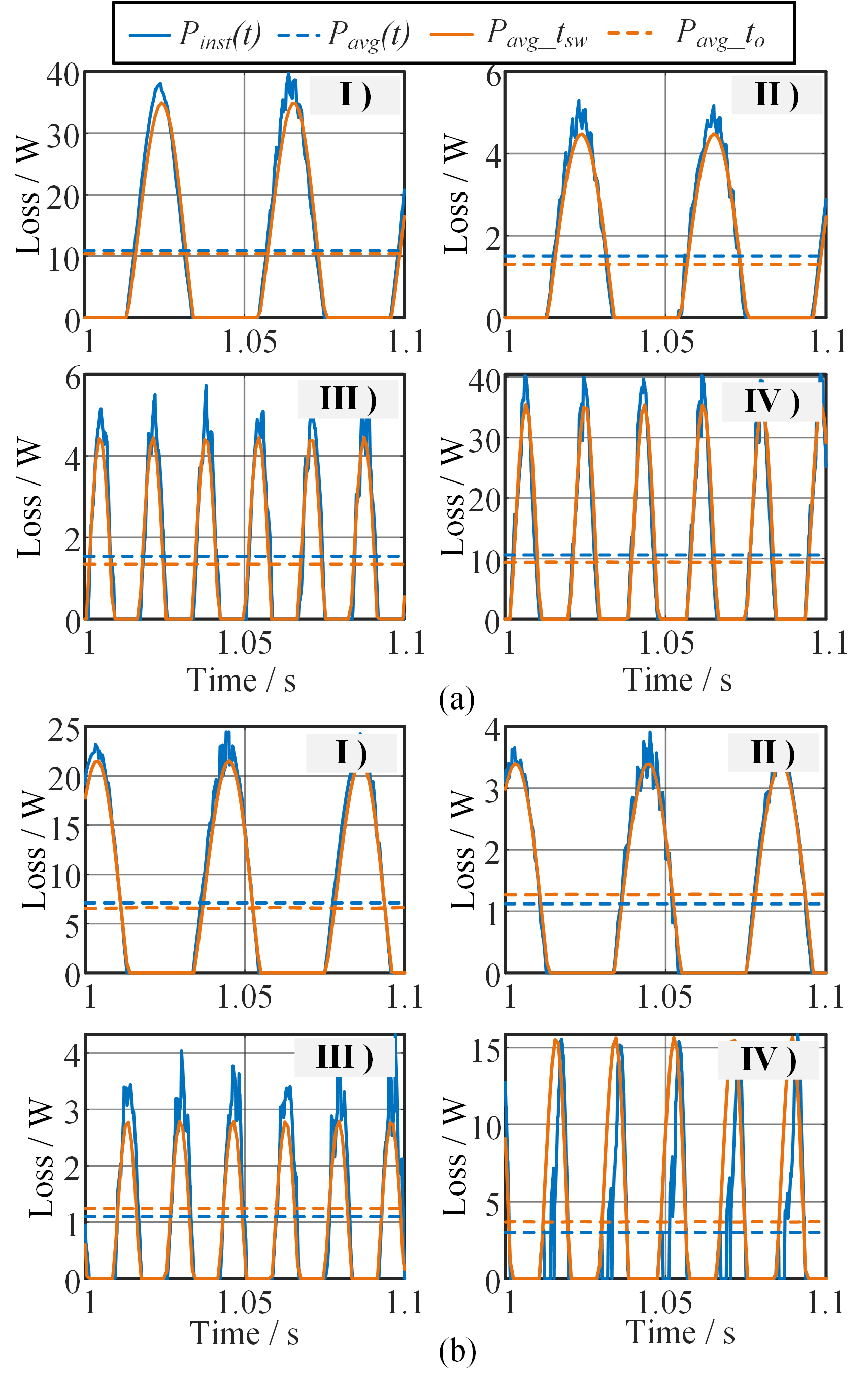}
	\caption{Loss models verification at four operation points under HWFET [solid line: average power loss using switching period average loss model (Benchmark is labeled as $P_\text{inst}(t)$, calculation result is labeled as $P_\text{avg}\_t_\text{sw}$), dashed line: average power loss using output period average loss model (Benchmark is labeled as $P_\text{avg}(t)$, calculation result is labeled as $P_\text{avg}\_t_\text{o}$)]. (a) IGBT loss. (b) Diode loss.}
    \label{fig_loss compare HWFET}
\end{figure}
\begin{figure}[t!]
	\centering
    \includegraphics[width=80mm]{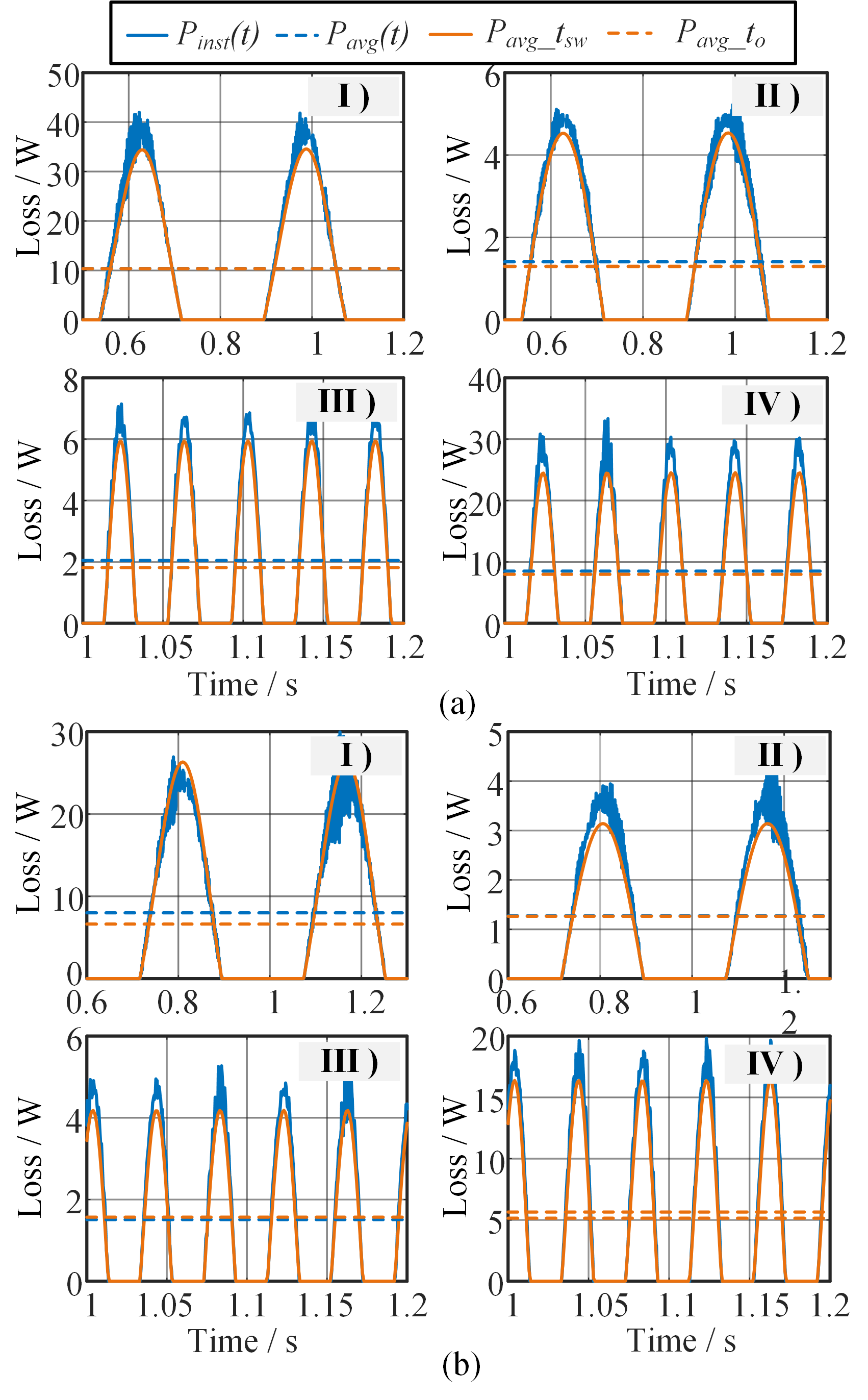}
	\caption{Loss models verification at four operation points under NYCC [solid line: average power loss using switching period average loss model (Benchmark is labeled as $P_\text{inst}(t)$, calculation result is labeled as $P_\text{avg}\_t_\text{sw}$), dashed line: average power loss using output period average loss model (Benchmark is labeled as $P_\text{avg}(t)$, calculation result is labeled as $P_\text{avg}\_t_\text{o}$)]. (a) IGBT loss. (b) Diode loss.}
    \label{fig_loss compare NYCC}
\end{figure}

Four specific operation points are selected from the aforementioned four different operation conditions to prove the loss models in steady state. The speed and torque at this four operation points are shown in Table.~\ref{Table:Operation points}. It should be noted that the torque is downscaled to meet the rated current of the power device. The power loss benchmark and the calculation results with different loss models under HWFET are shown in Fig.~\ref{fig_loss compare HWFET}, while the results under NYCC are shown in Fig.~\ref{fig_loss compare NYCC}. Note that in the following analysis, the estimated power loss and junction temperature using the output period average loss model are labeled as $t_\text{o}$, while the results using the switching period average loss model are labeled as $t_\text{sw}$. The maximum calculation error with different loss model is within 8.5~\%, which proves the effectiveness of using the two loss models to calculate power loss in steady state profiles. It can also be inferred from the results that when the vehicle is operating at different operation points, the power device loss has a significant difference. The power loss under situation I) and IV) is relatively large, which corresponding to the large torque. On the contrary, when the vehicle is driving with low torque, the power loss is small. Therefore, considering the highly dynamic driving cycles of the EV converter, the thermal loading of the power device would varies with different operation situations.

\begin{figure}[t!]
	\centering
    \includegraphics[width=80mm]{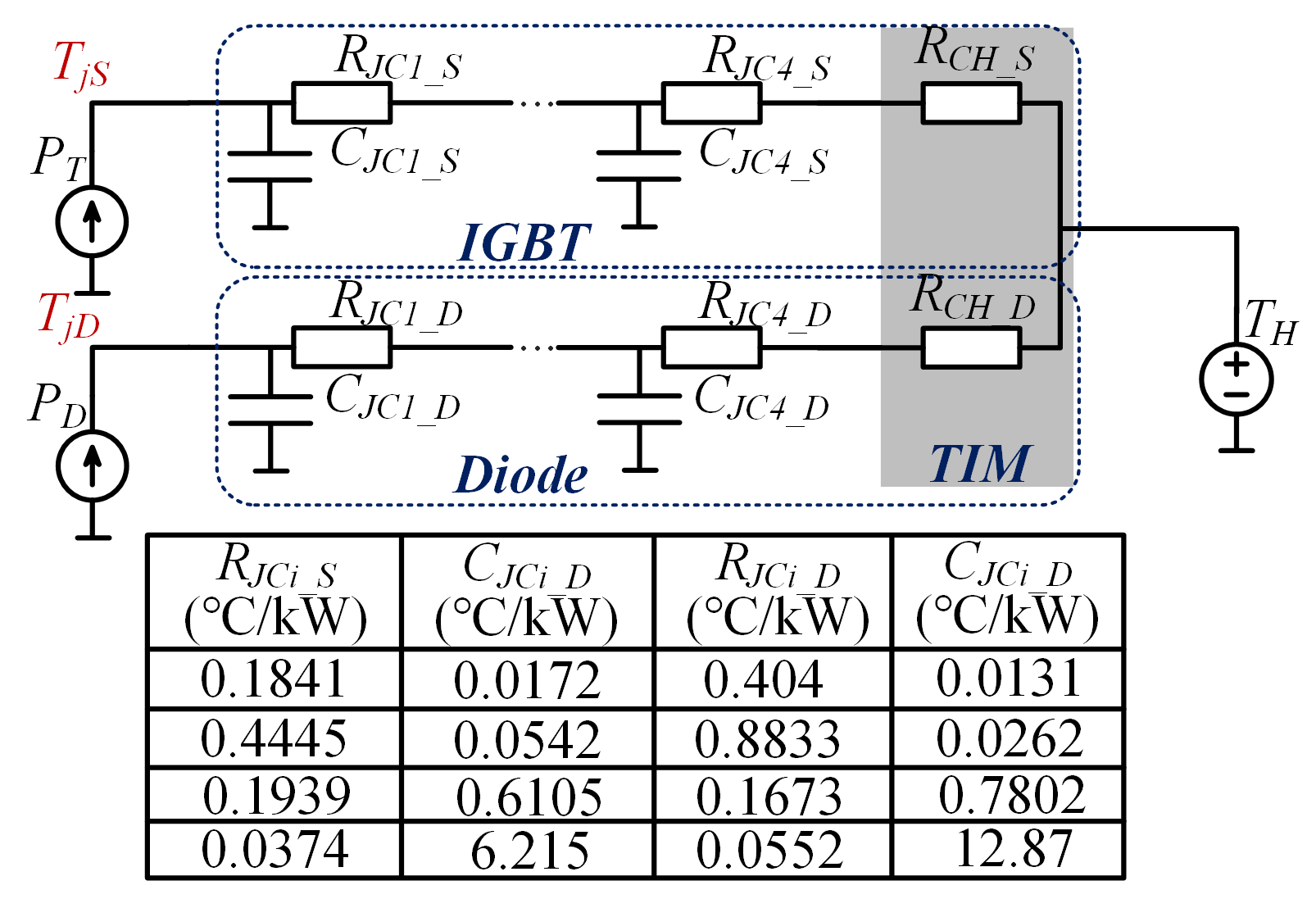}
	\caption{Four-order Cauer thermal model}
    \label{fig_thermal model}
\end{figure}

\begin{figure}[t!]
	\centering
    \includegraphics[width=80mm]{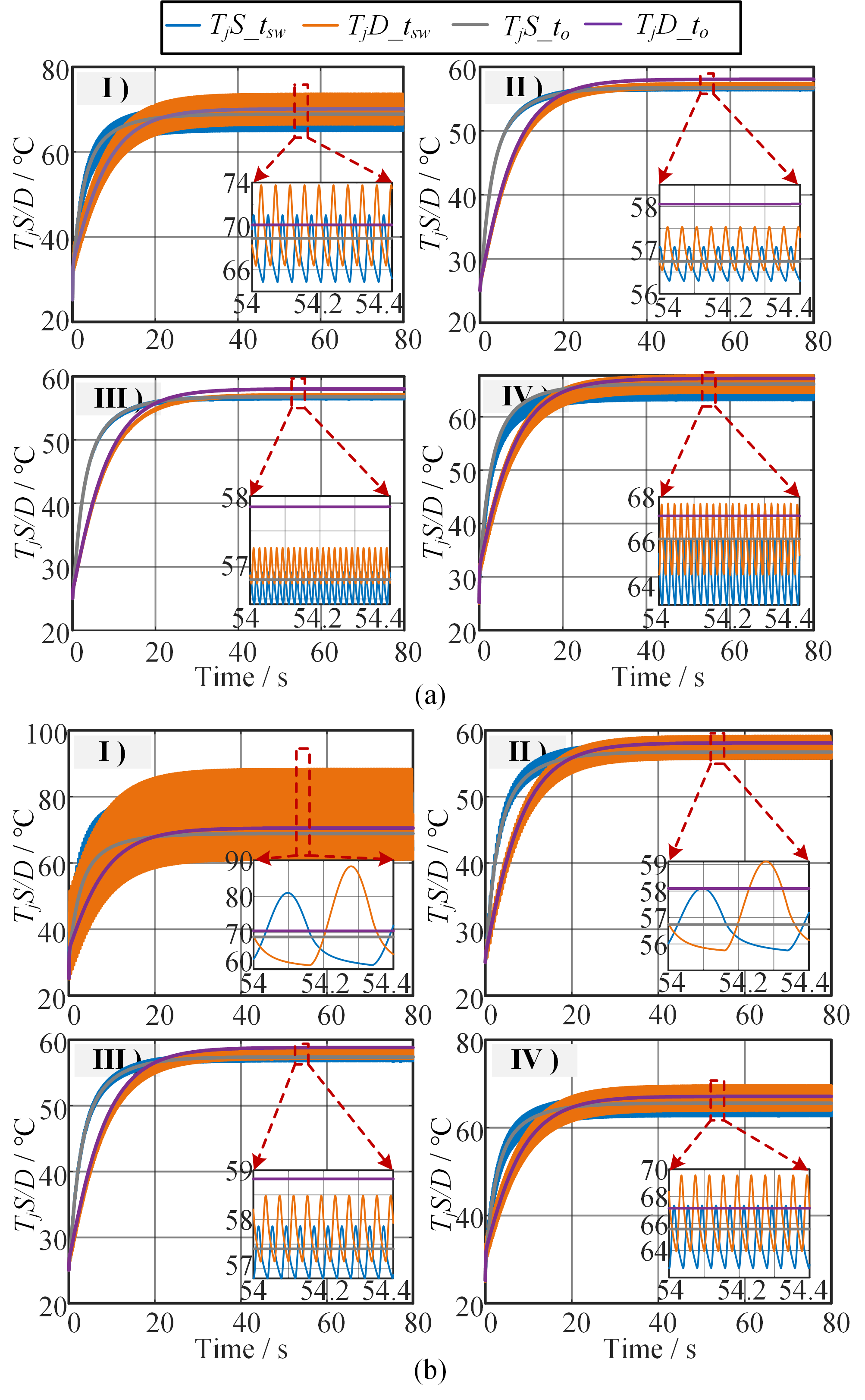}
	\caption{Estimated junction temperature under different driving cycles. (a) HWFET. (b) NYCC.}
    \label{fig_Tj compare}
\end{figure}

To study the effect of different loss models on the thermal loading under different driving cycles, the junction temperature of the power device is obtained with the four-order Cauer thermal model shown in Fig.~\ref{fig_thermal model}. $R_\text{JCi\_S}$ ($i$ = 1, 2, 3, 4) and $R_\text{JCi\_D}$ are the junction to case thermal resistance of the IGBT and diode, while $C_\text{JCi\_S}$ and $C_\text{JCi\_D}$ are the thermal capacitance. $R_\text{CH\_S}$ and $R_\text{CH\_D}$ are the case to heatsink thermal resistance of the IGBT and diode. $P_S$, $P_D$ denote the power loss of IGBT and diode, $T_\text{jS}$, $T_\text{jD}$ are the junction temperature of IGBT and diode. $T_H$ is the heatsink temperature and is set at 55 $^\circ$C. Accordingly, the estimated junction temperature are shown in Fig.~\ref{fig_Tj compare}.

The junction temperature estimation results shows significant effect of loss models and driving cycles. As can be seen from Fig.~\ref{fig_Tj compare}, when estimating the junction temperature with different loss models, the mean junction temperature $T_\text{jm}$ is almost the same, while the junction temperature swing $\Delta T_j$ in the output period is neglected when using the output period average loss model. Therefore, the effect of different loss models on the thermal loading estimation is reflected by the $\Delta T_j$. From Fig.~\ref{fig_Tj compare} (a), when the vehicle is driving in highway, the maximum $\Delta T_j$ is approximately 7 $^\circ$C at situation I), where the speed is low and torque is high. The minimum $\Delta T_j$ is approximately 0.5 $^\circ$C at situation III), where the speed is high and torque is low. However, from Fig.~\ref{fig_Tj compare}(b), when the vehicle is driving in city, the maximum $\Delta T_j$ can up to 27 $^\circ$C at situation I) and the minimum $\Delta T_j$ is still 1.5~$^\circ$C at situation III). Ref. \cite{Hirschmann2007} points out that the $\Delta T_j$ under 3 $^\circ$C has negligible effect on lifetime estimation. Therefore, compared with the small $\Delta T_j$ under the HWFET, using the output period averaging power loss model would bring significant error when estimating the lifetime of power devices in the NYCC.
\begin{figure}[t!]
	\centering
    \includegraphics[width=80mm]{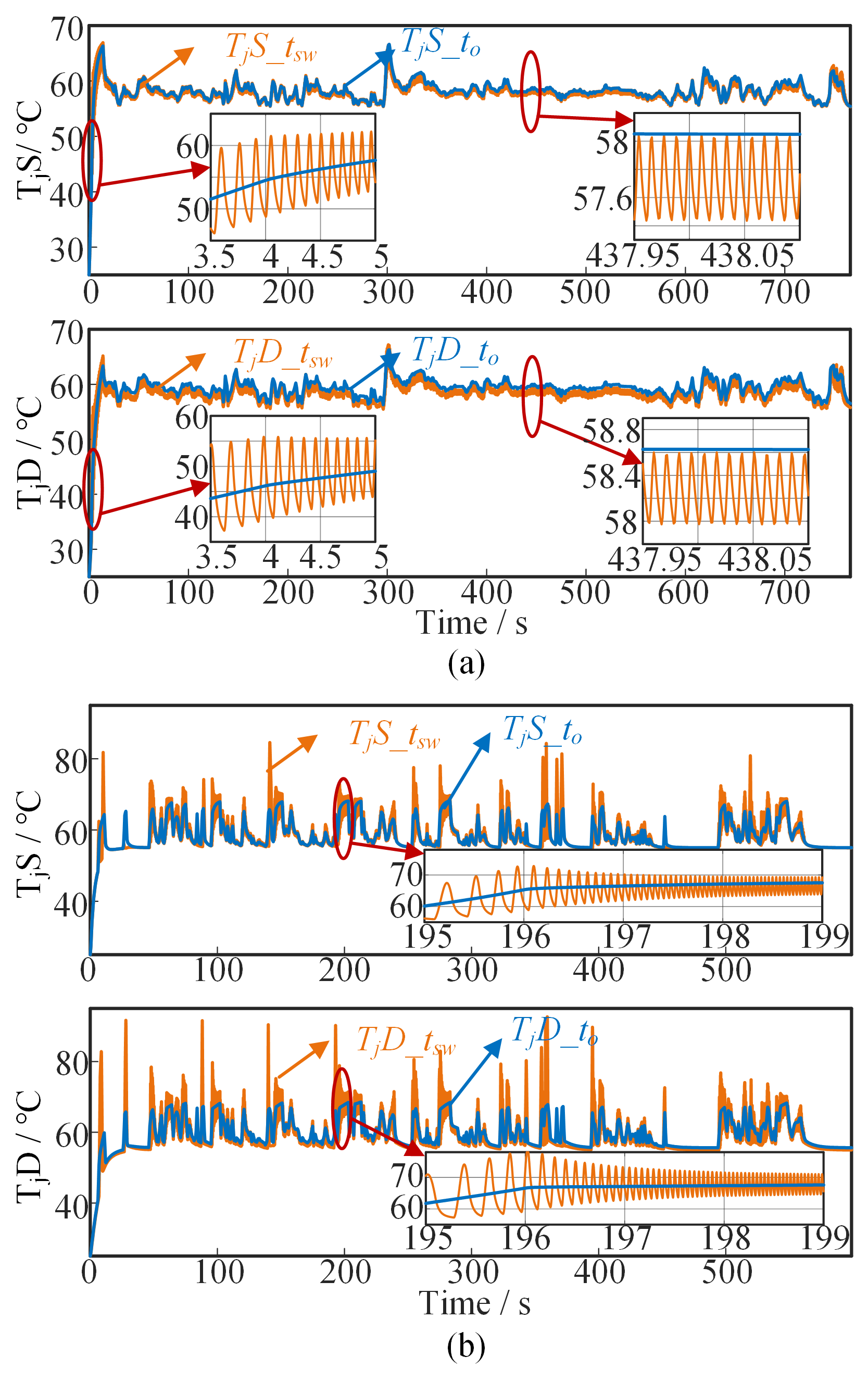}
	\caption{Estimated junction temperature with different loss model. (a) HWFET. (b) NYCC.}
    \label{fig_Tj}
\end{figure}

\section{Lifetime estimation with different power loss calculation methods}
Different from the aforementioned steady state based analysis, this section utilizes the dynamic driving cycles to illustrate the effect of loss model selection on the lifetime estimation of power devices in EV converters. The power device loss are calculated using different loss models under the HWFET and NYCC, the power loss is then fed into the thermal model in Fig.~\ref{fig_thermal model} to get the thermal loading. The thermal loading of the power device under different driving cycles are shown in Fig.~\ref{fig_Tj}.

From the results, the loss model selection has different impact on the estimated junction temperature under different driving cycles. The estimated junction temperature under the HWFET is shown in Fig.~\ref{fig_Tj}(a), and it shows different influence when the vehicle is accelerating and driving constantly at high speed. Under the acceleration period, the vehicle torque is large and the speed is low, which results in the large $\Delta T_j$ (approximately 20 $^\circ$C) in the power device, as shown in the zoomed window. However, under the constantly high speed driving period, the $\Delta T_j$ is very small (approximately 0.5 $^\circ$C) due to the low torque and high speed. It can be conclude that the estimation error is mostly induced by the acceleration period when using the fundamental period averaging loss model. Fortunately, we find that the high speed and low torque driving accounts for the most share in the HWFET in section II. Therefore, the influence of the loss model selection on the lifetime estimation could be acceptable under the HWFET. On the contrast, the estimated junction temperature under the NYCC driving cycle show extraordinary different results, as shown in Fig.~\ref{fig_Tj}(b). In the NYCC driving cycle, the low speed and high torque driving are of the most cases, thus the $\Delta T_j$ under the NYCC is very large (approximately 20 $^\circ$C in the zoomed window). The large $\Delta T_j$ is negelected when using the output period averaging loss model, which could bring large error in the lifetime estimation.

To illustrate the lifetime estimation error caused by different loss models, the obtained random thermal loading are translated to regular thermal cycles with the rainflow counting algorithm. The counting results is then used to estimate the lifetime of the power module. Based on the manufacturer provided data, the lifetime model for the selected IGBT module is 
\begin{equation}\label{equ:coffin manson}
    N_f = A \cdot (\Delta T_{j})^{\beta_1} \cdot \text{exp}(\frac{\beta_2}{T_\text{jmax}+273}) \cdot (\frac{t_\text{on}}{1.5})^{\beta_3}.
\end{equation}
where the $N_f$ is the number of cycles to failure. $T_\text{jmax}$ is the maximum junction temperature, $t_\text{on}$ is the turn-on time. $A$, $\beta_1$, $\beta_2$ and $\beta_3$ are fitting parameters provided by the manufacturer. Based on \cite{ludwig2010an2010} and \cite{Zhang2020}, $A$ = 1.42 $\times$ 10$^\text{12}$, $\beta_1 = -7.14$, $\beta_2 = 5154$ and $\beta_3 = -0.3$. $t_\text{on}$ has the limitation of 0.1~s~$< t_\text{on} <$~60~s.

With the counted cycles, the accumulated damage can be obtained according to the Palmgren-Miner linear cumulative damage law, and is expressed by
\begin{equation}\label{equ:Miner}
    D = \sum \frac{n_j}{N_{fj}}
\end{equation}
where $n_j$ is the cycles to a certain stress and $N_{fj}$ is the cycles to failure under the same stress. $D$ is the damage level of the device, where the device fails when $D$ equals to 1.

\begin{figure}[t!]
	\centering
    \includegraphics[width=80mm]{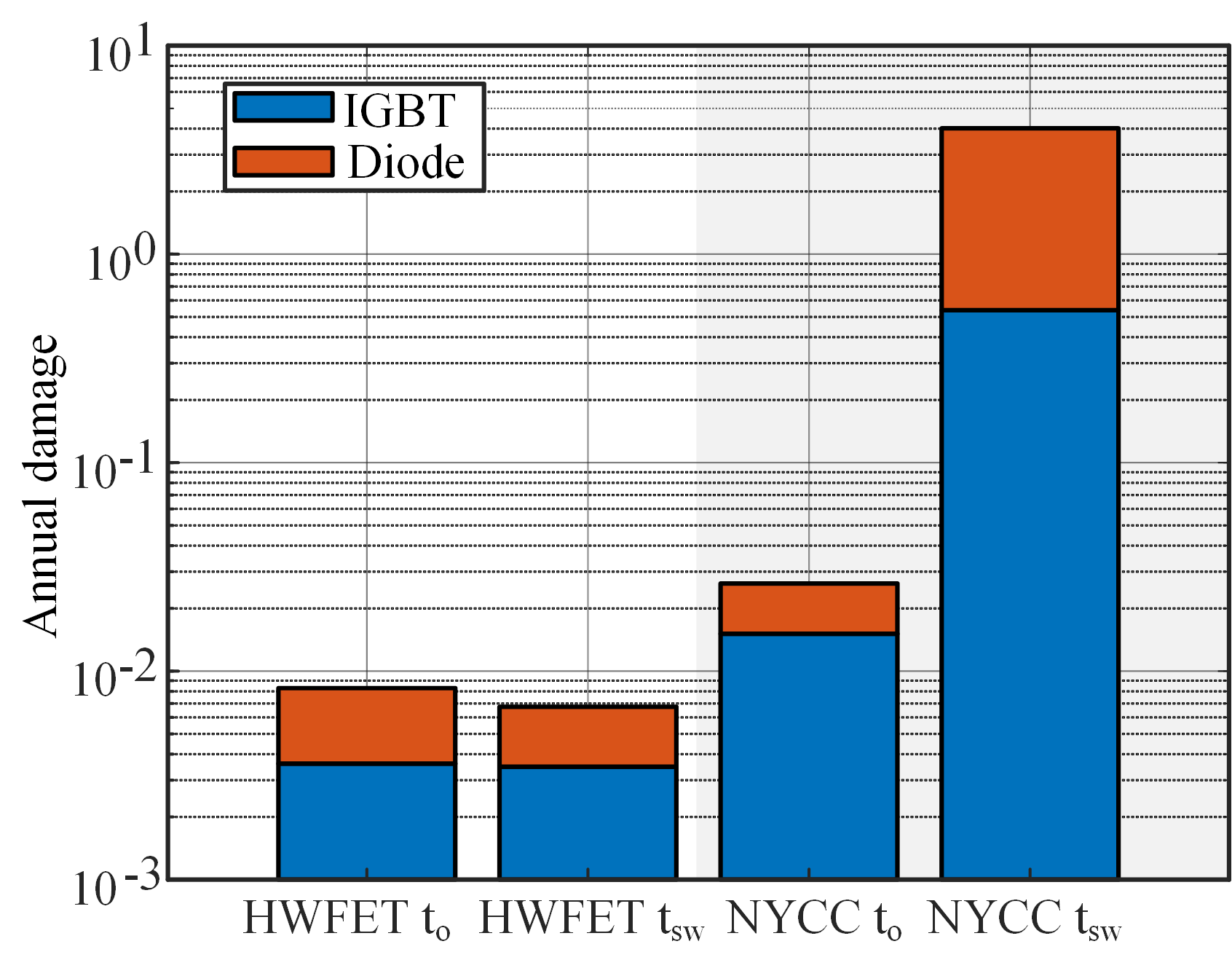}
	\caption{Annual damage of power device with different loss model.}
    \label{fig_damage}
\end{figure}
The annual damage results are shown in Fig.~\ref{fig_damage}, it can be inferred that the loss model selection strongly affects the estimated lifetime. Under the HWFET, the estimated annual damage with different loss models are very close. However, in the NYCC, the estimated damage using the switching period averaging model is much larger than that using the output period averaging model. Specifically, the estimated annual damage is 2.51 e$^{-2}$ for the IGBT and 1.12 e$^{-2}$ for the diode using the output period averaging model, while using the switching period averaging model it comes to 0.538 for the IGBT and 3.475 for the diode. The estimated damage using the output period averaging loss model is 35.8 times than using the switching period averaging loss model for IGBT, and 309.7 times for diode. Therefore, when estimating the IGBT module lifetime in EV converters, the loss model should be selected carefully depend on the driving cycle. For low speed and frequently stop-and-go city cycles, the switching cycle averaging loss model is more accurate. For high speed and low torque highway cycles, the output period averaging loss model can be chosen to get reasonable results while reducing the computational cost.

\section{Conclusion}
This paper studies the impact of different loss models on the IGBT module lifetime estimation in EV converters, where different driving cycles (highway and city) are considered. The effectiveness of two mostly used loss models, the output period averaging loss model and the switching period averaging loss model, are firstly verified with four different steady state operation points. However, the lifetime estimation results using the two loss models under dynamic driving cycles show extraordinary difference. Under the highway cycle, the estimated damage have a similar result. However, under the city cycle, the estimated lifetime by the output period averaging loss model is 35.8 times for IGBT and 309.7 times for diode higher than that applies the switching period averaging loss model. Therefore, when estimating the lifetime of IGBT modules in EV converters, the loss model should be carefully considered depending on the types of driving cycle.


\bibliographystyle{IEEEtran}
\bibliography{EV}

\end{document}